\documentclass{article}
\usepackage[a4paper, left=2cm, right=2cm, top=2.5cm, bottom=2.5cm]{geometry}
\usepackage{graphicx}
\usepackage{amsmath}
\usepackage{amssymb}
\usepackage{verbatim}
\usepackage{multirow}
\usepackage{tikz}
\usetikzlibrary{quantikz2}
\usepackage{float}
\usepackage{subcaption}
\usepackage{array}
\usepackage{pgfplots}
\pgfplotsset{compat=1.18}
\usepackage{amsthm}
\usepackage{url}
\usepackage{hyperref}
\usepackage{xcolor}
\hypersetup{colorlinks=false,
allbordercolors=white}
\usepackage{algorithmic}
\usepackage{algorithm}
\usepackage{authblk}

\usepackage{listings}
\usepackage{natbib}
\usepackage{amsmath,amssymb,amsfonts}
\usepackage{algorithmic}
\usepackage{listings}
\usepackage{graphicx}
\usepackage{textcomp}
\usepackage{xcolor}
\usepackage{courier}
\usepackage{multirow}
\usepackage{placeins}
\title{Triplet Loss Based Quantum Encoding for Class Separability}

\author{Marco Mordacci, Mahul Pandey, Paolo Santini and Michele Amoretti}

\affil{Quantum Software Laboratory, Department of Engineering and Architecture, University of Parma, 43124 Parma, Italy \\
\texttt{marco.mordacci1@unipr.it, mahul.pandey@unipr.it}}

\date{}
\begin{document}
\maketitle
\begin{abstract}
An efficient and data-driven encoding scheme is proposed to enhance the performance of variational quantum classifiers. This encoding is specially designed for complex datasets like images and seeks to help the classification task by producing input states that form well-separated clusters in the Hilbert space according to their classification labels. The encoding circuit is trained using a triplet loss function inspired by classical facial recognition algorithms, and class separability is measured via average trace distances between the encoded density matrices. Benchmark tests performed on various binary classification tasks on MNIST and MedMNIST datasets demonstrate considerable improvement over amplitude encoding with the same VQC structure while requiring a much lower circuit depth.
\end{abstract}



\section{Introduction}
\label{sec:intro}

The goal of quantum machine learning (QML)~\cite{biamonte2017quantum} is to take advantage of quantum computers to address learning tasks such as classification, regression, and content generation. A key component in QML is the data encoding strategy: the method by which classical data is embedded into the initial quantum states, to be subsequently manipulated using hybrid learning algorithms~\cite{schuld2018supervised, larose2020robust}. The encoding step is, in essence, the most important one, as it defines the working principle of the quantum algorithm, often constitutes the bottleneck step in terms of runtime, and the choice of encoding significantly influences the performance of the learning algorithm.

Commonly used encoding schemes, such as angle encoding and amplitude encoding \cite{schuld2018supervised}, are not efficient. Angle encoding maps classical data $x = (x_1, ..., x_n)$ into a quantum state using rotation angles:
\begin{equation}
    |x\rangle = \bigotimes_{i=1}^{n} \cos(x_i)|0\rangle + \sin(x_i)|1\rangle.
\end{equation}
However, angle encoding is not efficient in terms of the number of qubits, as it can encode only one feature per qubit.
Amplitude encoding, instead, maps a normalized classical $N$-dimensional data $x$ into the amplitudes of a quantum state:
\begin{equation}
    |\psi_x\rangle = \sum_{i=1}^N x_i |i\rangle
\end{equation}
where $N = 2^n$, $x_i$ is the $i^{th}$ element of $x$ and $|i\rangle$ is the $i^{th}$ computational basis state. However, amplitude encoding requires a quantum circuit whose depth increases exponentially with the number of qubits~\cite{mottonen2004transformation, sun2023asymptotically, belli2025srbb}. This is a relevant issue, as in the current Noisy Intermediate-Scale Quantum (NISQ) era, quantum devices are prone to noise and deeper circuits tend  to produce noise.
Furthermore, these methods do not take advantage of the structure of the data, which could be exploited for improved classification or generalization.

In this work, a novel approach to quantum encoding is explored, specifically aimed at classification problems, that is adaptive and data-driven. Instead of using an a-priori fixed encoding map, an encoding circuit is designed and optimized specifically for the classification task at hand. More precisely, it aims to embed input data in a way that explicitly achieves class separability in the Hilbert space. The method is inspired by FaceNet~\cite{Schroff2015}, a classical deep learning model for face recognition that employs a triplet loss function to encourage similar inputs to be mapped close together and dissimilar ones far apart in the embedding space.

The paper is organized as follows. In Section~\ref{sec:related}, the state-of-the-art is discussed and some of the recent developments in addressing the problem of quantum embeddings for machine learning are summarized. In Section~\ref{sec:problem}, the mathematical problem and the objective are defined. The strategy is described in Section~\ref{sec:solution}, where the triplet loss and how it drives the embedding circuit construction are explained. The basic variational classifier that uses these embedded states to make predictions is also described in that section. In Section~\ref{sec:evaluation}, results from both simulation and real hardware are reported. Finally, in Section~\ref{sec:conclusion} an overview of findings and future outlook is presented.

\section{Related Work}
\label{sec:related}
In recent years, the problem of encoding classical data into a quantum state has attracted considerable interest. A well-designed encoding scheme can simplify the task for the Variational Quantum Circuit (VQC), but the most commonly used methods, angle and amplitude encoding, present several challenges. Several algorithms have been developed in recent years to address this problem.

A comprehensive benchmarking of the most well-known encoding schemes, viz. amplitude, angle and IQP embedding~\cite{havlicek2018} has been carried out in~\cite{zang2025} on various datasets including the MNIST01 and MNIST08 subsets. These results are useful as a benchmark test for advanced encoding schemes that better take into account the data structure.

In recent years, a research field known as Quantum Architecture Search (QAS)~\cite{martyniuk2024quantum} has emerged, aiming to identify the optimal quantum circuit architecture for both the encoding stage and the VQC. Techniques such as Reinforcement Learning~\cite{kuo2021quantum, kolle2024optimizing}, evolutionary algorithms~\cite{jin2024practicality, wang2022quantumnas, altares2021automatic, sunkel2023ga4qco}, and particle swarm optimization were applied~\cite{kolle2024optimizing, mordacci2025training}.
Genetic algorithms~\cite{katoch2021review} are used to learn the whole architecture of the encoding scheme. Those are optimization techniques based on the theory of evolution. The algorithms evolve a population of individuals with the encoded feature maps, through the application of genetic operators. In each generation, the resulting offspring is selected in order to improve the objectives. In~\cite{altares2021automatic}, the authors optimized the quantum feature maps in a quantum kernel Support Vector Machine (SVM)~\cite{rebentrost2014quantum}. The genetic algorithm stores the circuit of each individual and aims to maximize the accuracy of the feature maps in modeling the data, while minimizing circuit complexity. The results show that it can produce a classifier with $100\%$ accuracy that generalizes well to unseen data.

In~\cite{lloyd2020quantum}, the authors, inspired by the so-called classical "metric learning"~\cite{bromley1993signature, chopra2005learning}, trained the quantum embedding using the trace and Hilbert-Schmidt distances.

Drawing inspiration from classical kernel methods~\cite{hofmann2008kernel}, quantum kernel methods map classical data into a high-dimensional quantum feature space~\cite{mercadier2023quantum, gentinetta2024complexity, yin2024experimental, ding2025quantum}. The similarity between data points is then calculated using a quantum kernel, which is essentially the inner product of the quantum states.
In~\cite{schuld2019quantum}, two strategies inspired by kernel theory~\cite{hofmann2008kernel} to find patterns in the data were proposed. One estimates intractable quantum kernels by feeding them into a classical kernel method; the other applies VQC to learn models that process the feature vectors.

Rath et al.~\cite{rath2024quantum} analyzed the impact of quantum encoding techniques (basis, superposition, angle, and amplitude encodings) in classical machine learning. Various classical algorithms were tested, such as SVM, Decision Tree, Random Forest, and AdaBoost.

In~\cite{gonzalez2024efficient}, an efficient method for amplitude encoding of real polynomial functions was proposed. 

Another interesting embedding is the Hamiltonian encoding~\cite{schuld2018supervised, di2021improving}, where classical data are encoded into the parameters of a system's Hamiltonian.

The choice of the encoding method plays a crucial role, as it can lead to the emergence of barren plateaus~\cite{larocca2025barren}, which cause the gradients to vanish exponentially with the size of the system, hindering effective training. The entanglement entropy~\cite{calabrese2004entanglement} generated by a quantum circuit can follow either a volume law or an area law. In a volume law regime, entanglement scales proportionally with the total number of qubits in the system; while in the area law scenario, entanglement entropy scales proportionally with the number of qubits at the boundary. Consequently, the applied encoding method can induce barren plateaus by generating excessive entanglement entropy~\cite{leone2024practical}. 
%\textit{Quantum embeddings for machine learning Lloyd-Schuld}, \textit{Quantum machine learning in feature Hilbert spaces, Schuld-Killoran}

\section{Problem Statement}
\label{sec:problem}

The goal of the data encoding strategy is to design an encoding that is well-suited for complex classification tasks involving unstructured data, such as image classification. As a concrete test case, a binary image classification task is considered using both MNIST and MedMNIST datasets~\cite{yang2021medmnist}. As a simple starting point, the datasets are restricted to the first two classes (e.g., “0” vs “1”). An embedding circuit $U_{emb}(\mathbf{x}^{(j)})$ is constructed (where ${\mathbf{x}^{(j)}}$ are the sets of feature vectors in the training dataset), such that the resulting state vectors $|\psi^{(j)}\rangle = U_{emb}(\mathbf{x}^{(j)}) |0\rangle$ lie in compact and mutually orthogonal subspaces of the Hilbert space depending on their class labels $y^{(j)}$. For an ideal encoding, the following should be obtained:

\begin{equation}
    \langle \psi^{(i)}|\psi^{(j)}\rangle \simeq 
\begin{cases}
1,\quad y^{(i)}=y^{(j)}\\
0,\quad y^{(i)}\neq y^{(j)}.
\end{cases}
\end{equation} 

Of course, an ideal encoding is not possible for such a complex, unstructured dataset; therefore, the algorithm aims to cluster the encoded state vectors into distinct regions of the Hilbert space. To introduce a natural metric on the space of states and to allow for future generalizations to noisy and partial measurements, density matrices $\rho^{(i)}$ are taken into account rather than state vectors. A natural metric on the Hilbert space is then the trace distance:
\begin{equation}
    D(\rho_1,\rho_2) = \frac{1}{2}\text{tr}\sqrt{(\rho_1-\rho_2)^\dagger(\rho_1-\rho_2)}.
\end{equation}

Two sets of pairs are defined:
\begin{itemize}
    \item Intra-class pairs $\mathcal{P}_{same} = \{(i,j)|y^{(i)} = y^{(j)}\}$,
    \item Inter-class pairs $\mathcal{P}_{diff} = \{(i,j)|y^{(i)} \neq y^{(j)}\}$.
\end{itemize}
The objective is to minimize the average distance $\mathbb{E}_{(i,j)\in\mathcal{P}_{same}}D(\rho^{(i)},\rho^{(j)})$ while maximizing the average distance $\mathbb{E}_{(i,j)\in\mathcal{P}_{diff}}D(\rho^{(i)},\rho^{(j)})$.

%%%%%%%%%%%%%%%%%%%%%%%%%%%%%%%
\section{Proposed Solution}
\label{sec:solution}

A triplet-loss-driven construction of a quantum encoding circuit that maps classical images to expressive quantum states is proposed. The entire algorithm proceeds in three main stages:
\begin{enumerate}
    \item Triplet selection,
    \item Greedy circuit construction,
    \item Variational classification circuit.
\end{enumerate}

\subsection{Triplet selection}

A strategy similar to the hard mining strategy in classical image recognition \cite{Schroff2015} is adopted. For each ordered pair of classes $(a,b)$, the following is chosen:
\begin{itemize}
    \item An anchor $A$: The median of all image vectors belonging to class $a$.
    \item A positive $P$: The image vector in class $a$ that is the farthest from the anchor (hard positive).
    \item A negative $N$: The image vector in class $b$ that is the closest to the anchor (hard negative).
\end{itemize}
The goal is to consider the worst-case scenario, selecting the farthest image of the same class as the anchor (positive) and the closest image from the other class (negative), in order to minimize the distance between the anchor and the positive, while maximizing the distance between the anchor and the negative.

\begin{figure}[!hbtp]
    \centering
    \includegraphics[width=0.5\linewidth]{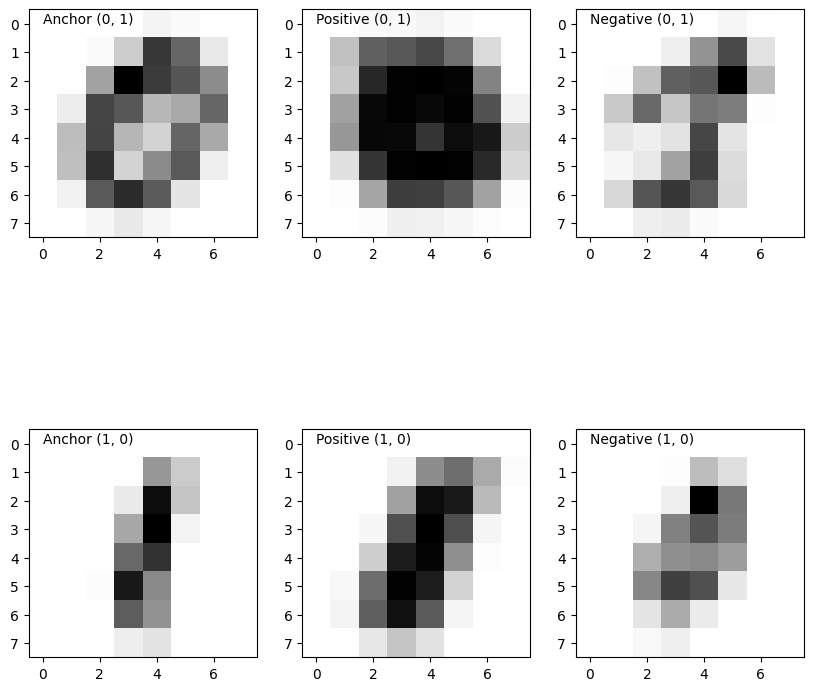}
    \caption{Triplets for the binary (0,1) subset of the MNIST dataset. The image resolution is $8\times 8$.}
    \label{fig:triplets}
\end{figure}

This triplet is designed to be maximally informative, promoting embeddings that tightly cluster class-relevant features while pushing away confusing examples. To demonstrate, the simple problem of binary classification with only digits "0" and "1" of the MNIST dataset is considered. There are only two triplets corresponding to $(0,1)$ and $(1,0)$ respectively. The triplets are shown in Figure~\ref{fig:triplets}.

\subsection{Greedy Circuit Construction}

Two hyperparameters are introduced:
\begin{itemize}
    \item A weight factor $w$ that introduces a relative weight between the positive and negative loss.
    \item A margin $m$ to adjust the degree of separation between the different clusters. 
\end{itemize}

The following objective function is minimized:
\begin{equation}
    J = \sum_{triplets} \max(0, w D(A,P)-D(A,N) + m).
\label{triploss}
\end{equation}

This is performed with a greedy combinatorial optimization algorithm:
\begin{enumerate}
    \item Start with an empty circuit.
    \item At each step $i$, pick the $i^{th}$ feature (pixel value) to encode.
    \item Incrementally add gates from a fixed pool: $\{R_X,R_Y,R_Z,CNOT,CZ\}$ at each step as follows:
    \begin{enumerate}
        \item Go through all possible gate-qubit combinations. For single-qubit rotations, the feature vector $x_i$ is encoded in the rotation angle.
        \item Temporarily append each combination to the previously constructed circuit, and evaluate the density matrices corresponding to each triplet element.
        \item Compute the triplet loss for each choice.
        \item Pick the gate-qubit combination that corresponds to the lowest triplet loss (greedy choice).
        \item If this best choice of gate turns out to be a $CNOT$ or a $CZ$, make sure to follow it up by a rotation gate that best encodes the $i^{th}$ feature. Thus, this feature is actually being encoded by an entangling gate followed by a rotation.
    \end{enumerate}
    \item Repeat this until all features have been encoded or a target depth is reached.
    \item Store the sequence of gate-qubit combinations: this serves as embedding circuit $U_{emb}$.
\end{enumerate}

The controlled rotations $(CR_x, CR_y, CR_z)$ are not included in the set of possible gates, as the optimal solutions that minimize the loss never select them. Instead, $CNOT$ gates and single-qubit rotations consistently perform better.

\subsection{Variational Classification Layer}

A trainable parameterized quantum circuit $U_{var}(\theta)$ is applied to the output of $U_{emb}(x)$, forming a hybrid quantum model. The parameters $\theta$ are optimized using a standard classical optimizer (Adam) to minimize the classification loss (cross-entropy).

\begin{figure}[hbtp!]
    \centering
    \includegraphics[width=0.5\linewidth]{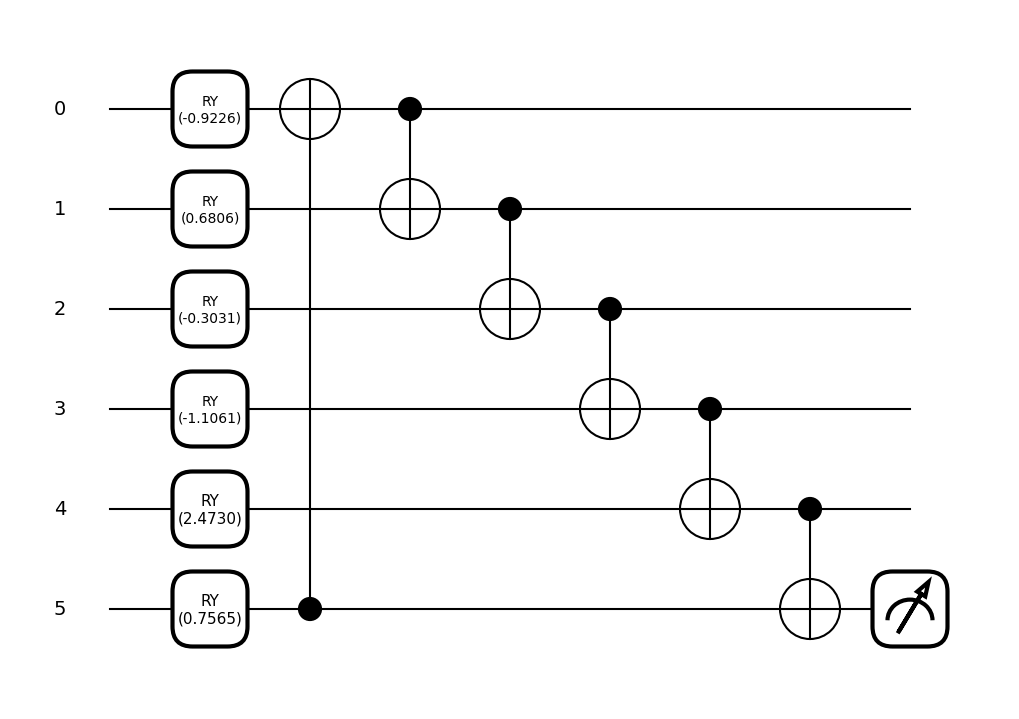}
    \caption{One layer of the variational circuit for a random choice of parameters.}
    \label{fig: vqc}
\end{figure}

The VQC of~\cite{zang2025} is used, each layer of which consists of:
\begin{itemize}
    \item A layer of $R_Y$ rotations, each parametrized by $\{\theta_1,...\theta_n\}$ applied to each qubit.
    \item A circular layer of CNOTs.
\end{itemize}
For a binary classification problem, the last qubit is measured and the probability of it being in the state $|1\rangle$ is obtained. One layer of the VQC is visualized in Fig. \ref{fig: vqc}.

Performance is evaluated using only one or two layers of the VQC, to demonstrate the improvement over the usual amplitude encoding scheme.

%%%%%%%%%%%%%%%%%%%%%%%%%%%%%%%
\section{Experimental Evaluation}
\label{sec:evaluation}

\subsection{Preprocessing}

Images are rescaled to $d\times d$ pixels, such that $n_{features} = d^2$. The proposed encoding scheme works better if no further rescaling is applied to the data set. 

% For now, we have restricted to a binary subset of our dataset, in particular the MNIST01 and MNIST08 subset. A straightforward extension to a multiclass classifier is being currently worked on.

\subsection{Hyperparameter Selection}

For a smaller scale of the problem, it is sufficient to set the margin parameter in (\ref{triploss}) to $m=0$. Also, for best results, it helps to choose a starting weight of $w = 1.0$ (which gives equal weightage to both cluster separation and condensation within the cluster), and dampening it by a small amount (0.01) at each step. The choice of using a cut-off of 0 for the loss function does not significantly affect the results, so the following objective is used:
\begin{equation}
    J = -\sum_{triplets} (D(A,N)-w D(A,P))
\end{equation}
with $w$ receiving a small damping at each iteration.

$N_{qubits}$ is set to $\log_2{n_{features}}$ since a lower number would lead to deeper circuits, worsen performance and noise resilience, and could lead to barren plateau problems.  

\begin{figure}[hbtp]
    \centering
    \includegraphics[width=1.0\linewidth]{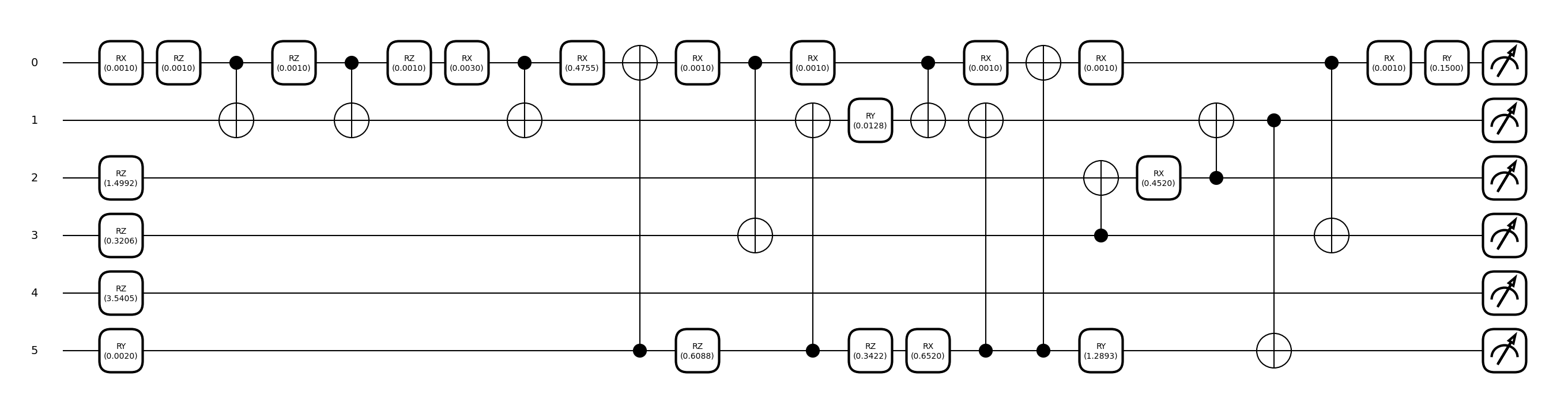}
    \caption{A typical embedding circuit. We have used the encoding of a 0 image as an example.}
    \label{fig:circuit6q}
\end{figure}

For the benefit of the reader, a typical embedding circuit generated for 8-by-8 MNIST01 images is presented in Figure~\ref{fig:circuit6q}, with the above selection of hyperparameters. The features are encoded in the rotation angles of the single-qubit rotation gates. Note that the total gate count in this circuit is less than 64, because consecutive rotations have been merged into a single rotation.
At the end of the circuit, all qubits are measured in order to retrieve the density matrices and compute the trace distances.
When the encoding circuit is ready, it will be followed by the VQC, without any measurement.

\subsection{MNIST Results}
\begin{figure}[hbtp!]
    \centering
    \includegraphics[width=1.0\linewidth]{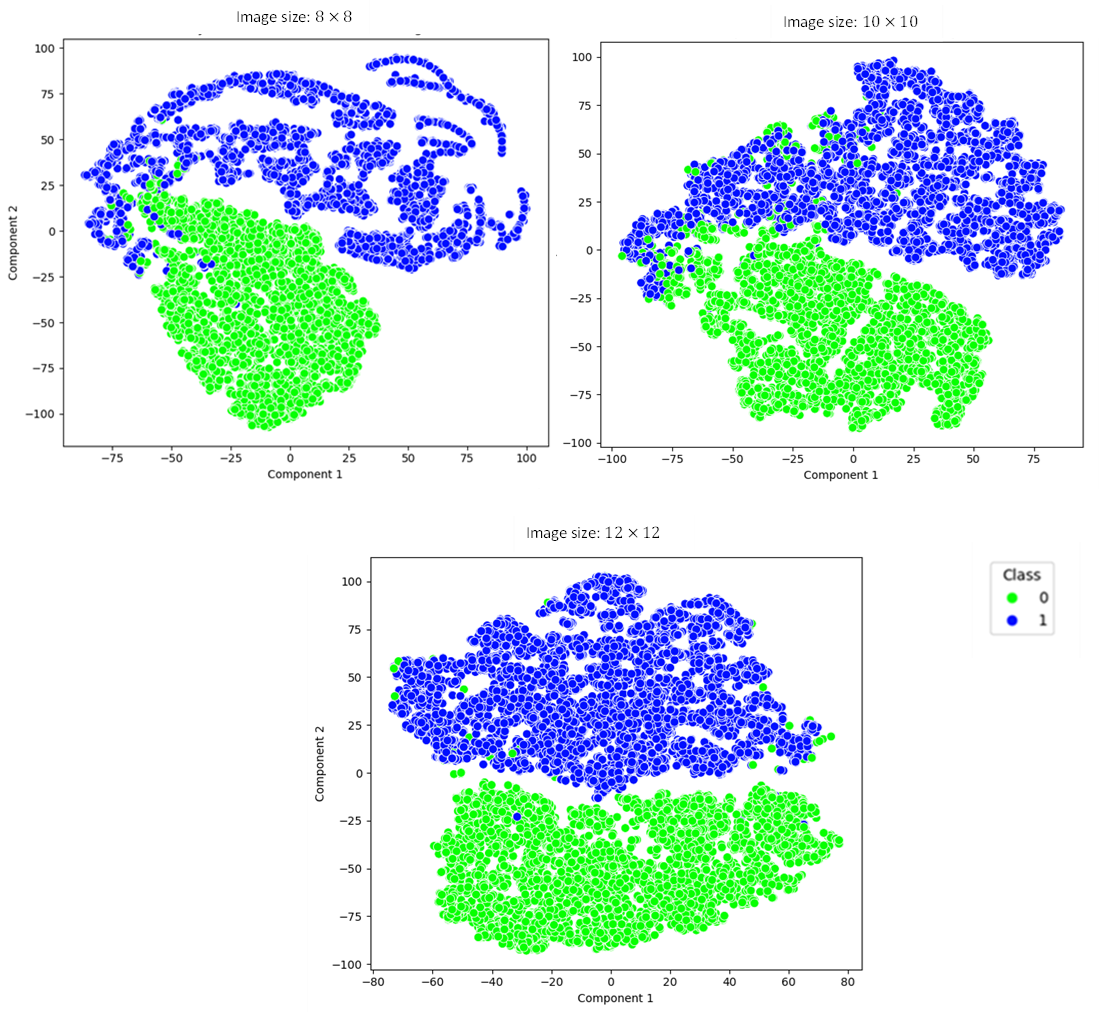}
    \caption{A visualization of the embedded density matrices using t-SNE, across different system sizes.}
    \label{fig:t-SNE}
\end{figure}

The clusters of the encoded density matrix are shown in Figure~\ref{fig:t-SNE}, with the help of a t-SNE embedding, for image resolutions ranging from 8-by-8 to 12-by-12 pixels. This demonstrates that the embedding performs quite well in cluster separation even with considerably low image resolutions.

\begin{figure}[hbtp!]
    \centering
    \includegraphics[width=1.0\linewidth]{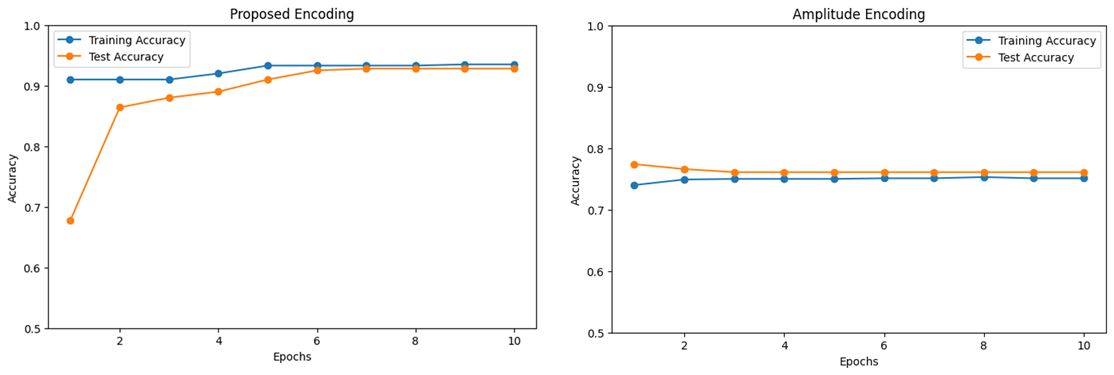}
    \caption{Epoch-wise accuracy curves during training of VQC with 1 layer, for $8\times 8$ MNIST images.}
    \label{fig:accuracy_curves}
\end{figure}

In Figure~\ref{fig:accuracy_curves}, the epoch-wise accuracy is plotted while training the VQC layers for 8-by-8 images (64 features), with the proposed encoding scheme and amplitude encoding, respectively. The proposed encoding matches or surpasses amplitude encoding in the accuracy of results for low resolution of images, while being more efficient in terms of circuit depth. 

The various metrics across feature size are shown in Table~\ref{tab:metricstable}. When a single layer of the VQC is used, the proposed method outperforms amplitude encoding (except in the 8-qubit case). However, with a more expressive two-layer VQC, the difference between the two encoding methods diminishes, and the choice of encoding circuit becomes less critical. Nonetheless, using fewer gates remains advantageous for execution on real devices, as it reduces noise, an effect that will be demonstrated in Section~\ref{lab:hw}.
In fact, the number of gates needed in the proposed method is much lower than those applied in amplitude encoding. Indeed, if the Mottonen state preparation~\cite{mottonen2004transformation} is applied to implement the amplitude embedding, the CNOT and rotation counts are:
\begin{itemize}
    \item 6 qubits: 228 CNOT, 251 Rotations,
    \item 7 qubits: 480 CNOT, 507 Rotations,
    \item 8 qubits: 980 CNOT, 1019 Rotations.
\end{itemize}
On the contrary, the total gate count of the proposed method is:
\begin{itemize}
    \item 64 features (6 qubits): 13 CNOT, 64 rotations at most (one for each feature, but consecutive rotations can be merged),
    \item 100 features (7 qubits): 14 CNOT, 100 rotations at most,
    \item 144 features (8 qubits): 24 CNOT, 144 rotations at most,
    \item 256 features (8 qubits): 47 CNOT, 256 rotations at most.
\end{itemize}
Note that amplitude encoding requires padding of the feature vector to match a power of 2, but the proposed method does not require such padding.

\begin{table}[hbtp!]
    \begin{tabular}{|>{\centering\arraybackslash}m{1cm}|>{\centering\arraybackslash}m{1cm}|>{\centering\arraybackslash}m{1.5cm}||>{\centering\arraybackslash}m{2cm}|>{\centering\arraybackslash}m{1.5cm}|>{\centering\arraybackslash}m{1cm}|>{\centering\arraybackslash}m{1.5cm}|>{\centering\arraybackslash}m{1cm}|>{\centering\arraybackslash}m{1cm}|}
        \hline
        $N_{features}$ & $N_{qubits}$ & $N_{layers}$ & Encoding & Accuracy & class & Precision & Recall & f1-score \\
        \hline
        \multirow{8}{*}{\centering $8 \times 8$} & \multirow{8}{*}{\centering 6} &  \multirow{4}{*}{\centering 1} & \multirow{2}{*}{\centering Amplitude} & \multirow{2}{*}{\centering $77\%$} & 0 & $0.67$ & $0.98$ & $0.80$ \\ 
        & & & & & 1 & $0.97$ & $0.59$ & $0.73$ \\
        \cline{4-9}
        & & & \multirow{2}{*}{\centering Proposed} & \multirow{2}{*}{\centering $93\%$} & 0 & $0.96$ & $0.89$ & $0.92$ \\ 
        & & & & & 1 & $0.91$ & $0.97$ & $0.94$ \\
        \cline{3-9}
        & & \multirow{4}{*}{\centering 2} & \multirow{2}{*}{\centering Amplitude} & \multirow{2}{*}{\centering $97\%$} & 0 & $0.97$ & $0.98$ & $0.97$ \\ 
        & & & & & 1 & $0.97$ & $0.96$ & $0.97$ \\
        \cline{4-9}
        & & & \multirow{2}{*}{\centering Proposed} & \multirow{2}{*}{\centering $97\%$} & 0 & $0.97$ & $0.96$ & $0.97$ \\ 
        & & & & & 1 & $0.97$ & $0.98$ & $0.97$ \\
        \cline{3-9}
        \hline

        \multirow{8}{*}{\centering $10 \times 10$} & \multirow{8}{*}{\centering 7} &  \multirow{4}{*}{\centering 1} & \multirow{2}{*}{\centering Amplitude} & \multirow{2}{*}{\centering $62\%$} & 0 & $0.58$ & $0.70$ & $0.63$ \\ 
        & & & & & 1 & $0.68$ & $0.56$ & $0.61$ \\
        \cline{4-9}
        & & & \multirow{2}{*}{\centering Proposed} & \multirow{2}{*}{\centering $91\%$} & 0 & $0.90$ & $0.91$ & $0.91$ \\ 
        & & & & & 1 & $0.92$ & $0.91$ & $0.92$ \\
        \cline{3-9}
        & & \multirow{4}{*}{\centering 2} & \multirow{2}{*}{\centering Amplitude} & \multirow{2}{*}{\centering $87\%$} & 0 & $0.78$ & $0.99$ & $0.87$ \\ 
        & & & & & 1 & $0.99$ & $0.76$ & $0.86$ \\
        \cline{4-9}
        & & & \multirow{2}{*}{\centering Proposed} & \multirow{2}{*}{\centering $92\%$} & 0 & $0.93$ & $0.90$ & $0.91$ \\ 
        & & & & & 1 & $0.90$ & $0.94$ & $0.92$ \\
        \cline{3-9}
        \hline

        \multirow{8}{*}{\centering $12 \times 12$} & \multirow{8}{*}{\centering 8} &  \multirow{4}{*}{\centering 1} & \multirow{2}{*}{\centering Amplitude} & \multirow{2}{*}{\centering $94\%$} & 0 & $0.90$ & $0.98$ & $0.94$ \\ 
        & & & & & 1 & $0.98$ & $0.91$ & $0.94$ \\
        \cline{4-9}
        & & & \multirow{2}{*}{\centering Proposed} & \multirow{2}{*}{\centering $97\%$} & 0 & $0.96$ & $0.96$ & $0.96$ \\ 
        & & & & & 1 & $0.97$ & $0.97$ & $0.97$ \\
        \cline{3-9}
        & & \multirow{4}{*}{\centering 2} & \multirow{2}{*}{\centering Amplitude} & \multirow{2}{*}{\centering $97.5\%$} & 0 & $0.97$ & $0.97$ & $0.97$ \\ 
        & & & & & 1 & $0.98$ & $0.97$ & $0.97$ \\
        \cline{4-9}
        & & & \multirow{2}{*}{\centering Proposed} & \multirow{2}{*}{\centering $97\%$} & 0 & $0.98$ & $0.96$ & $0.97$ \\ 
        & & & & & 1 & $0.97$ & $0.98$ & $0.97$ \\
        \cline{3-9}
        \hline

        \multirow{8}{*}{\centering $16 \times 16$} & \multirow{8}{*}{\centering 8} &  \multirow{4}{*}{\centering 1} & \multirow{2}{*}{\centering Amplitude} & \multirow{2}{*}{\centering $95\%$} & 0 & $0.90$ & $1.0$ & $0.95$ \\ 
        & & & & & 1 & $1.0$ & $0.91$ & $0.94$ \\
        \cline{4-9}
        & & & \multirow{2}{*}{\centering Proposed} & \multirow{2}{*}{\centering $54\%$} & 0 & $0.50$ & $0.74$ & $0.60$ \\ 
        & & & & & 1 & $0.62$ & $0.36$ & $0.46$ \\
        \cline{3-9}
        & & \multirow{4}{*}{\centering 2} & \multirow{2}{*}{\centering Amplitude} & \multirow{2}{*}{\centering $95\%$} & 0 & $0.95$ & $0.94$ & $0.95$ \\ 
        & & & & & 1 & $0.94$ & $0.95$ & $0.94$ \\
        \cline{4-9}
        & & & \multirow{2}{*}{\centering Proposed} & \multirow{2}{*}{\centering $84\%$} & 0 & $0.86$ & $0.79$ & $0.82$ \\ 
        & & & & & 1 & $0.83$ & $0.89$ & $0.86$ \\
        \cline{3-9}
        \hline
    \end{tabular}
\caption{Comparison of metrics for different feature size}
\label{tab:metricstable} 
\end{table}

\subsection{MedMNIST Results}

In this case, a learning rate of $0.01$ and a batch size of $32$ are used. For each result, the training is executed 5 times and the mean results are taken. For the Tissue and OCT datasets, a smaller training set of 3500 samples (1750 samples for each class) is used to reduce the encoding circuit generation time.
The results obtained with the proposed method are compared to those achieved using amplitude encoding. Only one layer of the circuit shown in Fig. \ref{fig: vqc} is used.

Table~\ref{tab:medmnistTest8} reports the test set accuracies on various MedMNIST datasets, where the images are rescaled to $8  \times 8$. The class-wise precision, recall, and f1-score of a random execution are presented.
An improvement in performance cannot be observed. In fact, both methods tend to predict a single class for all examples, indicating a failure to learn meaningful patterns. The only improvement is visible in the organA dataset; while amplitude encoding still predicts a single class, the proposed method begins to learn.

\begin{table}[hbtp!]
    \begin{tabular}{|>{\centering\arraybackslash}m{1cm}|>{\centering\arraybackslash}m{1cm}|>{\centering\arraybackslash}m{1.5cm}||>{\centering\arraybackslash}m{2cm}|>{\centering\arraybackslash}m{1.5cm}|>{\centering\arraybackslash}m{1cm}|>{\centering\arraybackslash}m{1.5cm}|>{\centering\arraybackslash}m{1cm}|>{\centering\arraybackslash}m{1cm}|}
        \hline
        $N_{features}$ & $N_{qubits}$ & Dataset & Encoding & Accuracy & class & Precision & Recall & f1-score \\
        \hline
        \multirow{32}{*}{\centering $8 \times 8$} & \multirow{32}{*}{\centering 6} & \multirow{4}{*}{\centering Chest} & \multirow{2}{*}{\centering Amplitude} & \multirow{2}{*}{\centering $54\%$} & 0 & $0.54$ & $0.65$ & $0.59$ \\ 
        & & & & & 1 & $0.49$ & $0.38$ & $0.43$ \\
        \cline{4-9}
        & & & \multirow{2}{*}{\centering Proposed} & \multirow{2}{*}{\centering $53\%$} & 0 & $0.53$ & $1.0$ & $0.69$ \\ 
        & & & & & 1 & $0.55$ & $0.001$ & $0.002$ \\
        \cline{3-9}
        & & \multirow{4}{*}{\centering Breast} & \multirow{2}{*}{\centering Amplitude} & \multirow{2}{*}{\centering $73\%$} & 0 & $0.0$ & $0.0$ & $0.0$ \\ 
        & & & & & 1 & $0.73$ & $1.0$ & $0.84$ \\
        \cline{4-9}
        & & & \multirow{2}{*}{\centering Proposed} & \multirow{2}{*}{\centering $73\%$} & 0 & $0.0$ & $0.0$ & $0.0$ \\ 
        & & & & & 1 & $0.73$ & $1.0$ & $0.84$ \\
        \cline{3-9}
        & & \multirow{4}{*}{\centering oct} & \multirow{2}{*}{\centering Amplitude} & \multirow{2}{*}{\centering $55\%$} & 0 & $0.59$ & $0.32$ & $0.41$ \\ 
        & & & & & 1 & $0.53$ & $0.78$ & $0.64$ \\
        \cline{4-9}
        & & & \multirow{2}{*}{\centering Proposed} & \multirow{2}{*}{\centering $50\%$} & 0 & $0.0$ & $0.0$ & $0.0$ \\ 
        & & & & & 1 & $0.50$ & $1.0$ & $0.67$ \\
        \cline{3-9}
        & & \multirow{4}{*}{\centering Tissue} & \multirow{2}{*}{\centering Amplitude} & \multirow{2}{*}{\centering $44\%$} & 0 & $0.89$ & $0.39$ & $0.54$ \\ 
        & & & & & 1 & $0.14$ & $0.67$ & $0.23$ \\
        \cline{4-9}
        & & & \multirow{2}{*}{\centering Proposed} & \multirow{2}{*}{\centering $87\%$} & 0 & $0.87$ & $1.0$ & $0.93$ \\ 
        & & & & & 1 & $0.0$ & $0.0$ & $0.0$\\
        \cline{3-9}
        & & \multirow{4}{*}{\centering Pneumonia} & \multirow{2}{*}{\centering Amplitude} & \multirow{2}{*}{\centering $62\%$} & 0 & $0.0$ & $0.0$ & $0.0$ \\ 
        & & & & & 1 & $0.62$ & $1.0$ & $0.77$ \\
        \cline{4-9}
        & & & \multirow{2}{*}{\centering Proposed} & \multirow{2}{*}{\centering $62\%$} & 0 & $0.0$ & $0.0$ & $0.0$ \\ 
        & & & & & 1 & $0.62$ & $1.0$ & $0.77$ \\
        \cline{3-9}
        & & \multirow{4}{*}{\centering OrganA} & \multirow{2}{*}{\centering Amplitude} & \multirow{2}{*}{\centering $58.5\%$} & 0 & $0.58$ & $1.0$ & $0.72$ \\ 
        & & & & & 1 & $0.0$ & $0.0$ & $0.0$ \\
        \cline{4-9}
        & & & \multirow{2}{*}{\centering Proposed} & \multirow{2}{*}{\centering \boldmath$67\%$} & 0 & \boldmath$0.78$ & \boldmath$0.63$ & \boldmath$0.70$  \\ 
        & & & & & 1 & \boldmath$0.61$ & \boldmath$0.76$ & \boldmath$0.68$ \\
        \cline{3-9}
        & & \multirow{4}{*}{\centering OrganC} & \multirow{2}{*}{\centering Amplitude} & \multirow{2}{*}{\centering $66\%$} & 0 & $0.66$ & $1.0$ & $0.79$ \\ 
        & & & & & 1 & $0.0$ & $0.0$ & $0.0$ \\
        \cline{4-9}
        & & & \multirow{2}{*}{\centering Proposed} & \multirow{2}{*}{\centering $66\%$} & 0 & $0.66$ & $1.0$ & $0.79$  \\ 
        & & & & & 1 & $0.0$ & $0.0$ & $0.0$\\
        \cline{3-9}
        & & \multirow{4}{*}{\centering OrganS} & \multirow{2}{*}{\centering Amplitude} & \multirow{2}{*}{\centering $65\%$} & 0 & $0.65$ & $1.0$ & $0.79$ \\ 
        & & & & & 1 & $0.0$ & $0.0$ & $0.0$ \\
        \cline{4-9}
        & & & \multirow{2}{*}{\centering Proposed} & \multirow{2}{*}{\centering $65\%$} & 0 & $0.65$ & $1.0$ & $0.79$  \\ 
        & & & & & 1 & $0.0$ & $0.0$ & $0.0$\\
        \cline{3-9}
        \hline
    \end{tabular}
    \caption{\centering $8 \times 8$ MedMNIST results on the test set with different dataset.}
    \label{tab:medmnistTest8}   
\end{table}

However, with $16 \times 16$ images, the differences between the two encoding methods become visible, as shown in Table~\ref{tab:medmnistTest16}. Indeed, for Chest, Breast, and OrganC datasets, the proposed encoding method performs similarly to the amplitude encoding in terms of accuracy; on the other hand, amplitude encoding performs better with the Tissue and OCT datasets, due to the fact that the smaller training set loses the true member of the triplet for each class. However, for the Tissue dataset, the proposed encoding method starts recognizing some members of the positive class, even though in the test set there are more negative than positive examples. Furthermore, the proposed method achieves better accuracy with the Pneumonia, OrganA, and OrganS datasets. In this case, while amplitude encoding does not learn meaningful patterns and still predicts the same class every time (such as in the $8 \times 8$ scenario), the proposed encoder is starting to learn both classes.

\begin{table}[hbtp!]
    \begin{tabular}{|>{\centering\arraybackslash}m{1cm}|>{\centering\arraybackslash}m{1cm}|>{\centering\arraybackslash}m{1.5cm}||>{\centering\arraybackslash}m{2cm}|>{\centering\arraybackslash}m{1.5cm}|>{\centering\arraybackslash}m{1cm}|>{\centering\arraybackslash}m{1.5cm}|>{\centering\arraybackslash}m{1cm}|>{\centering\arraybackslash}m{1cm}|}
        \hline
        $N_{features}$ & $N_{qubits}$ & Dataset & Encoding & Accuracy & Class & Precision & Recall & f1-score \\
        \hline
        \multirow{32}{*}{\centering $16 \times 16$} & \multirow{32}{*}{\centering 8} & \multirow{4}{*}{\centering Chest} & \multirow{2}{*}{\centering Amplitude} & \multirow{2}{*}{\centering $52.5\%$} & 0 & $0.55$ & $0.63$ & $0.59$ \\ 
        & & & & & 1 & $0.50$ & $0.41$ & $0.45$ \\
        \cline{4-9}
        & & & \multirow{2}{*}{\centering Proposed} & \multirow{2}{*}{\centering $52\%$} & 0 & $0.55$ & $0.64$ & $0.59$ \\ 
        & & & & & 1 & $0.49$ & $0.40$ & $0.44$ \\
        \cline{3-9}
        & & \multirow{4}{*}{\centering Breast} & \multirow{2}{*}{\centering Amplitude} & \multirow{2}{*}{\centering $62\%$} & 0 & $0.38$ & $0.71$ & $0.50$ \\ 
        & & & & & 1 & $0.85$ & $0.58$ & $0.69$ \\
        \cline{4-9}
        & & & \multirow{2}{*}{\centering Proposed} & \multirow{2}{*}{\centering $53\%$} & 0 & $0.30$ & $0.57$ & $0.40$  \\ 
        & & & & & 1 & $0.77$ & $0.52$ & $0.62$ \\
        \cline{3-9}
        & & \multirow{4}{*}{\centering oct} & \multirow{2}{*}{\centering Amplitude} & \multirow{2}{*}{\centering $54\%$} & 0 & $0.54$ & $0.57$ & $0.56$ \\ 
        & & & & & 1 & $0.54$ & $0.51$ & $0.53$ \\
        \cline{4-9}
        & & & \multirow{2}{*}{\centering Proposed} & \multirow{2}{*}{\centering $50\%$} & 0 & $0.75$ & $0.01$ & $0.02$ \\ 
        & & & & & 1 & $0.50$ & $1.00$ & $0.67$\\
        \cline{3-9}
        & & \multirow{4}{*}{\centering Tissue} & \multirow{2}{*}{\centering Amplitude} & \multirow{2}{*}{\centering $66\%$} & 0 & $0.87$ & $0.71$ & $0.78$ \\ 
        & & & & & 1 & $0.12$ & $0.27$ & $0.17$ \\
        \cline{4-9}
        & & & \multirow{2}{*}{\centering Proposed} & \multirow{2}{*}{\centering $82\%$} & 0 & $0.87$ & $0.93$ & $0.90$ \\ 
        & & & & & 1 & $0.11$ & $0.05$ & $0.07$ \\
        \cline{3-9}
        & & \multirow{4}{*}{\centering Pneumonia} & \multirow{2}{*}{\centering Amplitude} & \multirow{2}{*}{\centering $62\%$} & 0 & $0.0$ & $0.0$ & $0.0$ \\ 
        & & & & & 1 & $0.62$ & $1.0$ & $0.77$ \\
        \cline{4-9}
        & & & \multirow{2}{*}{\centering Proposed} & \multirow{2}{*}{\centering \boldmath$59\%$} & 0 & \boldmath$0.46$ & \boldmath$0.76$ & \boldmath$0.57$ \\ 
        & & & & & 1 & \boldmath$0.76$ & \boldmath$0.47$ & \boldmath$0.58$ \\
        \cline{3-9}
        & & \multirow{4}{*}{\centering OrganA} & \multirow{2}{*}{\centering Amplitude} & \multirow{2}{*}{\centering $58.5\%$} & 0 & $0.58$ & $1.0$ & $0.72$ \\ 
        & & & & & 1 & $0.0$ & $0.0$ & $0.0$ \\
        \cline{4-9}
        & & & \multirow{2}{*}{\centering Proposed} & \multirow{2}{*}{\centering \boldmath$60\%$} & 0 & \boldmath$0.6$ & \boldmath$0.90$ & \boldmath$0.72$  \\ 
        & & & & & 1 & \boldmath$0.6$ & \boldmath$0.19$ & \boldmath$0.29$ \\
        \cline{3-9}
        & & \multirow{4}{*}{\centering OrganC} & \multirow{2}{*}{\centering Amplitude} & \multirow{2}{*}{\centering $66\%$} & 0 & $0.66$ & $1.0$ & $0.79$ \\ 
        & & & & & 1 & $0.0$ & $0.0$ & $0.0$ \\
        \cline{4-9}
        & & & \multirow{2}{*}{\centering Proposed} & \multirow{2}{*}{\centering $66\%$} & 0 & $0.66$ & $1.0$ & $0.79$  \\ 
        & & & & & 1 & $0.0$ & $0.0$ & $0.0$\\
        \cline{3-9}
        & & \multirow{4}{*}{\centering OrganS} & \multirow{2}{*}{\centering Amplitude} & \multirow{2}{*}{\centering $65\%$} & 0 & $0.65$ & $1.0$ & $0.79$ \\ 
        & & & & & 1 & $0.0$ & $0.0$ & $0.0$ \\
        \cline{4-9}
        & & & \multirow{2}{*}{\centering Proposed} & \multirow{2}{*}{\centering \boldmath$58\%$} & 0 & \boldmath$0.64$ & \boldmath$0.88$ & \boldmath$0.74$  \\ 
        & & & & & 1 & \boldmath$0.32$ & \boldmath$0.10$ & \boldmath$0.16$\\
        \cline{3-9}
        \hline
    \end{tabular}
    \caption{\centering $16 \times 16$ MedMNIST results on the test set with different dataset.}
    \label{tab:medmnistTest16}   
\end{table}

With $28 \times 28$ images; the proposed encoding shows a larger improvement in recognizing both classes, as shown in Table~\ref{tab:medmnistTest28}. Now, it achieves better results also on the OrganC and the Tissue datasets and, in general, gives a more balanced f1-score across classes than the amplitude encoding.

\begin{table}[hbtp!]
    \begin{tabular}{|>{\centering\arraybackslash}m{1cm}|>{\centering\arraybackslash}m{1cm}|>{\centering\arraybackslash}m{1.5cm}||>{\centering\arraybackslash}m{2cm}|>{\centering\arraybackslash}m{1.5cm}|>{\centering\arraybackslash}m{1cm}|>{\centering\arraybackslash}m{1.5cm}|>{\centering\arraybackslash}m{1cm}|>{\centering\arraybackslash}m{1cm}|}
        \hline
        $N_{features}$ & $N_{qubits}$ & Dataset & Encoding & Accuracy & Class & Precision & Recall & f1-score \\
        \hline
        \multirow{32}{*}{\centering $28 \times 28$} & \multirow{32}{*}{\centering 10} & \multirow{4}{*}{\centering Chest} & \multirow{2}{*}{\centering Amplitude} & \multirow{2}{*}{\centering $54\%$} & 0 & $0.54$ & $0.96$ & $0.69$ \\ 
        & & & & & 1 & $0.60$ & $0.06$ & $0.11$ \\
        \cline{4-9}
        & & & \multirow{2}{*}{\centering Proposed} & \multirow{2}{*}{\centering $52\%$} & 0 & $0.54$ & $0.68$ & $0.60$ \\ 
        & & & & & 1 & $0.48$ & $0.33$ & $0.40$ \\
        \cline{3-9}
        & & \multirow{4}{*}{\centering Breast} & \multirow{2}{*}{\centering Amplitude} & \multirow{2}{*}{\centering $49\%$} & 0 & $0.24$ & $0.43$ & $0.31$ \\ 
        & & & & & 1 & $0.71$ & $0.51$ & $0.59$ \\
        \cline{4-9}
        & & & \multirow{2}{*}{\centering Proposed} & \multirow{2}{*}{\centering $46\%$} & 0 & $0.19$ & $0.31$ & $0.24$ \\ 
        & & & & & 1 & $0.67$ & $0.52$ & $0.59$ \\
        \cline{3-9}
        & & \multirow{4}{*}{\centering oct} & \multirow{2}{*}{\centering Amplitude} & \multirow{2}{*}{\centering $51\%$} & 0 & $0.52$ & $0.37$ & $0.43$ \\ 
        & & & & & 1 & $0.51$ & $0.65$ & $0.57$ \\
        \cline{4-9}
        & & & \multirow{2}{*}{\centering Proposed} & \multirow{2}{*}{\centering $50\%$} & 0 & $0.50$ & $0.48$ & $0.49$ \\ 
        & & & & & 1 & $0.50$ & $0.52$ & $0.51$ \\
        \cline{3-9}
        & & \multirow{4}{*}{\centering Tissue} & \multirow{2}{*}{\centering Amplitude} & \multirow{2}{*}{\centering $38\%$} & 0 & $0.88$ & $0.34$ & $0.49$ \\ 
        & & & & & 1 & $0.13$ & $0.69$ & $0.22$ \\
        \cline{4-9}
        & & & \multirow{2}{*}{\centering Proposed} & \multirow{2}{*}{\centering \boldmath$68\%$} & 0 & \boldmath$0.87$ & \boldmath$0.75$ & \boldmath$0.80$ \\ 
        & & & & & 1 & \boldmath$0.12$ & \boldmath$0.23$ & \boldmath$0.15$ \\
        \cline{3-9}
        & & \multirow{4}{*}{\centering Pneumonia} & \multirow{2}{*}{\centering Amplitude} & \multirow{2}{*}{\centering $62\%$} & 0 & $0.0$ & $0.0$ & $0.0$ \\ 
        & & & & & 1 & $0.62$ & $1.0$ & $0.77$ \\
        \cline{4-9}
        & & & \multirow{2}{*}{\centering Proposed} & \multirow{2}{*}{\centering \boldmath$55\%$} & 0 & \boldmath$0.42$ & \boldmath$0.44$ & \boldmath$0.43$ \\ 
        & & & & & 1 & \boldmath$0.66$ & \boldmath$0.64$ & \boldmath$0.65$ \\
        \cline{3-9}
        & & \multirow{4}{*}{\centering OrganA} & \multirow{2}{*}{\centering Amplitude} & \multirow{2}{*}{\centering $58.5\%$} & 0 & $0.58$ & $1.0$ & $0.72$ \\ 
        & & & & & 1 & $0.0$ & $0.0$ & $0.0$ \\
        \cline{4-9}
        & & & \multirow{2}{*}{\centering Proposed} & \multirow{2}{*}{\centering \boldmath$63\%$} & 0 & \boldmath$0.66$ & \boldmath$0.74$ & \boldmath$0.69$  \\ 
        & & & & & 1 & \boldmath$0.58$ & \boldmath$0.49$ & \boldmath$0.53$ \\
        \cline{3-9}
        & & \multirow{4}{*}{\centering OrganC} & \multirow{2}{*}{\centering Amplitude} & \multirow{2}{*}{\centering $66\%$} & 0 & $0.66$ & $1.0$ & $0.79$ \\ 
        & & & & & 1 & $0.0$ & $0.0$ & $0.0$ \\
        \cline{4-9}
        & & & \multirow{2}{*}{\centering Proposed} & \multirow{2}{*}{\centering \boldmath$54\%$} & 0 & \boldmath$0.69$ & \boldmath$0.53$ & \boldmath$0.60$  \\ 
        & & & & & 1 & \boldmath$0.37$ & \boldmath$0.54$ & \boldmath$0.44$\\
        \cline{3-9}
        & & \multirow{4}{*}{\centering OrganS} & \multirow{2}{*}{\centering Amplitude} & \multirow{2}{*}{\centering $65\%$} & 0 & $0.65$ & $1.0$ & $0.79$ \\ 
        & & & & & 1 & $0.0$ & $0.0$ & $0.0$ \\
        \cline{4-9}
        & & & \multirow{2}{*}{\centering Proposed} & \multirow{2}{*}{\centering \boldmath$58\%$} & 0 & \boldmath$0.70$ & \boldmath$0.65$ & \boldmath$0.67$  \\ 
        & & & & & 1 & \boldmath$0.43$ & \boldmath$0.49$ & \boldmath$0.46$\\
        \cline{3-9}
        \hline
    \end{tabular}
    \caption{\centering $28 \times 28$ MedMNIST results on the test set with different dataset.}
    \label{tab:medmnistTest28}   
\end{table}

\begin{figure}[hbtp!]
    \centering
    \includegraphics[width=0.5\linewidth]{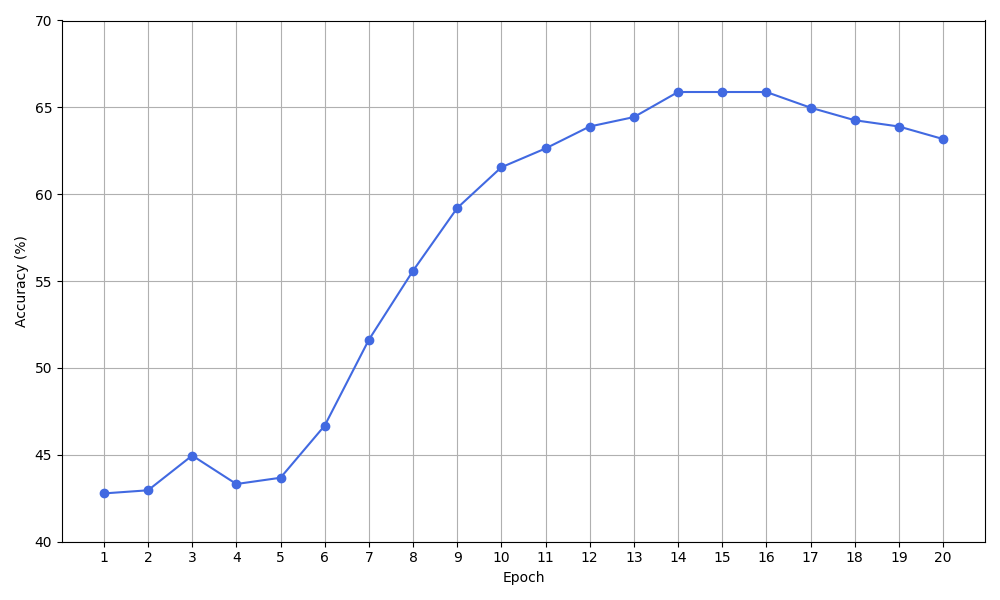}
    \caption{Accuracy on $8 \times 8$ organA validation dataset during training.}
    \label{fig:validation}
\end{figure}

In Figure~\ref{fig:validation}, an example of training on the OrganA validation set is presented.

Furthermore, the MNIST dataset is trained with only 3500 samples to understand how the smaller part of the training set affects the results for the Tissue and OCT datasets. The proposed method achieves an accuracy of $75\%$, while the amplitude encoding reaches approximately $80\%$. These results show a performance deterioration due to the absence of the original anchor, positive, and negative images in the reduced training set. Therefore, other techniques will be explored to reduce the computational time.

\subsection{Real Hardware Execution}
\label{lab:hw}
The produced encoding circuit with the trained VQC for $8 \times 8$ MNIST images is tested on Rigetti Ankaa-3 and IQM Garnet quantum computers, through AWS Braket. Each circuit is executed with 1024 shots. The first 500 samples of the test set are taken into account. In Table~\ref{tab:rigetti} the accuracy, precision, recall, and f1-score are reported. On Rigetti Ankaa-3, the results are worse due to noise, as the encoding circuit still has a high depth, even though the number of rotations and CNOT gates is lower than in the amplitude encoding. However, on IQM Garnet, the noise does not affect the results. The accuracy is identical to that obtained in simulation; there are only minor variations in precision, recall, and f1-score.

\begin{table}[hbtp!]
    \begin{tabular} {|c|c|c|c|c|c|c|c|}
        \hline
        $n_{features}$ & $N_{qubits}$ & Hardware & Accuracy & Class & Precision & Recall & f1-score\\ 
        \hline
         \multirow{6}{*}{\centering $8 \times 8$} & \multirow{6}{*}{\centering 6} & \multirow{2}{*}{\centering Rigetti Ankaa-3} & \multirow{2}{*}{\centering $50\%$} & 0 & 0.33 & 0.14 & 0.19 \\ 
         & & & & 1 & 0.78 & 0.54 & 0.64\\
         \cline{3-8}
         & & \multirow{2}{*}{\centering IQM Garnet} & \multirow{2}{*}{\centering $96.8\%$} & 0 & 0.97 & 0.95 & 0.96 \\
         & & & & 1 & 0.97 & 0.98 & 0.97 \\
         \cline{3-8}
         \cline{3-8}
         & & \multirow{2}{*}{\centering Simulation} & \multirow{2}{*}{\centering $96.8\%$} & 0 & 0.99 & 0.93 & 0.96 \\
         & & & & 1 & 0.95 & 0.99 & 0.97 \\
        \hline
    \end{tabular} \\

    \caption{Comparison of accuracies on simulation and real hardware during the testing of the MNIST dataset.}
    \label{tab:rigetti}   
\end{table}

%%%%%%%%%%%%%%%%%%%%%%%%%%%%%%%
\section{Conclusion}
\label{sec:conclusion}
In this work, a novel quantum encoding scheme based on triplet loss is proposed. Specifically, the gates applied to the encoding circuit are selected to maximize the distance between different classes and minimize the distance between samples of the same class. The generated encoding circuit has a lower depth than amplitude encoding while using the same number of features, and it does not lose any information, unlike angle encoding, which uses only a few features to encode the data.

This encoding scheme is tested on various binary classification tasks. The MNIST dataset and various MedMNIST datasets are taken into account to evaluate the encoding performance. Furthermore, the method is also evaluated on real quantum hardware to determine the impact of noise on the results.

The proposed encoding scheme is very successful in class separation for simple and distinctive image datasets like the MNIST dataset, and for low-resolution images achieves better results than amplitude encoding with a much lesser circuit depth. For more complex image datasets like the MedMNIST, where variations between images of different classes are much harder to spot, the encoding scheme performs more poorly when low-resolution images are used; however, at high enough resolutions, it starts to capture these inter-class differences better.

Future work will involve modifying the encoding generator to further enhance performance. In particular, the current "hard" selection of triplets is too simplistic, and more complex image datasets are likely to benefit from a more effective triplet-mining strategy. Additionally, the "greedy" algorithm for adding gates may be refined to incorporate strategies that better capture correlations among different features and to reduce the encoding circuit generation time, as computational time increases with higher-resolution images or larger datasets.

Currently, when different gate configurations applied to different qubits yield the same optimal loss, the gates acting on the first qubits are preferentially selected. As a result, a disproportionately high number of operations is applied to the first qubits, leading to an increased circuit depth, and making the other qubits redundant. This issue will be addressed in the future.

Furthermore, an evaluation of deeper VQC will be performed, and the algorithm will also be tested in multiclass scenarios.

\section*{Acknowledgment}
We acknowledge the financial support from Spoke 10 - ICSC - ``National Research Centre in High Performance Computing, Big Data and Quantum Computing'', funded by European Union – NextGenerationEU.
This research benefits from the High Performance Computing facility of the University of Parma, Italy (HPC.unipr.it).

%%
%% Define the bibliography file to be used
\bibliographystyle{unsrt}
\bibliography{bibliography}

\begin{thebibliography}{10}

\bibitem{biamonte2017quantum}
Jacob Biamonte, Peter Wittek, Nicola Pancotti, Patrick Rebentrost, Nathan Wiebe, and Seth Lloyd.
\newblock Quantum machine learning.
\newblock {\em Nature}, 549(7671):195--202, 2017.

\bibitem{schuld2018supervised}
Maria Schuld and Francesco Petruccione.
\newblock {\em Supervised learning with quantum computers}, volume~17.
\newblock Springer, 2018.

\bibitem{larose2020robust}
Ryan LaRose and Brian Coyle.
\newblock Robust data encodings for quantum classifiers.
\newblock {\em Physical Review A}, 102(3):032420, 2020.

\bibitem{mottonen2004transformation}
Mikko Mottonen, Juha~J Vartiainen, Ville Bergholm, and Martti~M Salomaa.
\newblock Transformation of quantum states using uniformly controlled rotations.
\newblock {\em arXiv preprint quant-ph/0407010}, 2004.

\bibitem{sun2023asymptotically}
Xiaoming Sun, Guojing Tian, Shuai Yang, Pei Yuan, and Shengyu Zhang.
\newblock Asymptotically optimal circuit depth for quantum state preparation and general unitary synthesis.
\newblock {\em IEEE Transactions on Computer-Aided Design of Integrated Circuits and Systems}, 42(10):3301--3314, 2023.

\bibitem{belli2025srbb}
Giacomo Belli, Marco Mordacci, and Michele Amoretti.
\newblock Srbb-based quantum state preparation.
\newblock In {\em Proceedings of the 22nd ACM International Conference on Computing Frontiers}, pages 172--175, 2025.

\bibitem{Schroff2015}
Florian Schroff, Dmitry Kalenichenko, and James Philbin.
\newblock Facenet: A unified embedding for face recognition and clustering.
\newblock In {\em 2015 IEEE Conference on Computer Vision and Pattern Recognition (CVPR)}, pages 815--823, 2015.

\bibitem{havlicek2018}
Vojtěch Havlíček, Antonio~D. Córcoles, Kristan Temme, Aram~W. Harrow, Abhinav Kandala, Jerry~M. Chow, and Jay~M. Gambetta.
\newblock Supervised learning with quantum-enhanced feature spaces.
\newblock {\em Nature}, 567(7747):209–212, March 2019.

\bibitem{zang2025}
Orlane Zang, Grégoire Barrué, and Tony Quertier.
\newblock Benchmarking data encoding methods in quantum machine learning, 2025.

\bibitem{martyniuk2024quantum}
Darya Martyniuk, Johannes Jung, and Adrian Paschke.
\newblock Quantum architecture search: a survey.
\newblock In {\em 2024 IEEE International Conference on Quantum Computing and Engineering (QCE)}, volume~1, pages 1695--1706. IEEE, 2024.

\bibitem{kuo2021quantum}
En-Jui Kuo, Yao-Lung~L Fang, and Samuel Yen-Chi Chen.
\newblock Quantum architecture search via deep reinforcement learning.
\newblock {\em arXiv preprint arXiv:2104.07715}, 2021.

\bibitem{kolle2024optimizing}
Michael K{\"o}lle, Daniel Seidl, Maximilian Zorn, Philipp Altmann, Jonas Stein, and Thomas Gabor.
\newblock Optimizing variational quantum circuits using metaheuristic strategies in reinforcement learning.
\newblock In {\em 2024 IEEE International Conference on Quantum Computing and Engineering (QCE)}, volume~2, pages 323--328. IEEE, 2024.

\bibitem{jin2024practicality}
Alex Jin, Tarun Dutta, Manas Mukherjee, Jos{\'e} Latorre, et~al.
\newblock Practicality of training a quantum machine in the nisq era.
\newblock In {\em APS Division of Atomic, Molecular and Optical Physics Meeting Abstracts}, volume 2024, pages D00--081, 2024.

\bibitem{wang2022quantumnas}
Hanrui Wang, Yongshan Ding, Jiaqi Gu, Yujun Lin, David~Z Pan, Frederic~T Chong, and Song Han.
\newblock Quantumnas: Noise-adaptive search for robust quantum circuits.
\newblock In {\em 2022 IEEE International Symposium on High-Performance Computer Architecture (HPCA)}, pages 692--708. IEEE, 2022.

\bibitem{altares2021automatic}
Sergio Altares-L{\'o}pez, Angela Ribeiro, and Juan~Jos{\'e} Garc{\'\i}a-Ripoll.
\newblock Automatic design of quantum feature maps.
\newblock {\em Quantum Science and Technology}, 6(4):045015, 2021.

\bibitem{sunkel2023ga4qco}
Leo S{\"u}nkel, Darya Martyniuk, Denny Mattern, Johannes Jung, and Adrian Paschke.
\newblock Ga4qco: genetic algorithm for quantum circuit optimization.
\newblock {\em arXiv preprint arXiv:2302.01303}, 2023.

\bibitem{mordacci2025training}
Marco Mordacci and Michele Amoretti.
\newblock Training variational quantum circuits using particle swarm optimization.
\newblock In {\em 2025 IEEE International Conference on Quantum Computing and Engineering (QCE)}. IEEE, 2025.

\bibitem{katoch2021review}
Sourabh Katoch, Sumit~Singh Chauhan, and Vijay Kumar.
\newblock A review on genetic algorithm: past, present, and future.
\newblock {\em Multimedia tools and applications}, 80:8091--8126, 2021.

\bibitem{rebentrost2014quantum}
Patrick Rebentrost, Masoud Mohseni, and Seth Lloyd.
\newblock Quantum support vector machine for big data classification.
\newblock {\em Physical review letters}, 113(13):130503, 2014.

\bibitem{lloyd2020quantum}
Seth Lloyd, Maria Schuld, Aroosa Ijaz, Josh Izaac, and Nathan Killoran.
\newblock Quantum embeddings for machine learning.
\newblock {\em arXiv preprint arXiv:2001.03622}, 2020.

\bibitem{bromley1993signature}
Jane Bromley, Isabelle Guyon, Yann LeCun, Eduard S{\"a}ckinger, and Roopak Shah.
\newblock Signature verification using a" siamese" time delay neural network.
\newblock {\em Advances in neural information processing systems}, 6, 1993.

\bibitem{chopra2005learning}
Sumit Chopra, Raia Hadsell, and Yann LeCun.
\newblock Learning a similarity metric discriminatively, with application to face verification.
\newblock In {\em 2005 IEEE computer society conference on computer vision and pattern recognition (CVPR'05)}, volume~1, pages 539--546. IEEE, 2005.

\bibitem{hofmann2008kernel}
Thomas Hofmann, Bernhard Sch{\"o}lkopf, and Alexander~J Smola.
\newblock Kernel methods in machine learning.
\newblock 2008.

\bibitem{mercadier2023quantum}
Mathieu Mercadier.
\newblock Quantum-enhanced versus classical support vector machine: An application to stock index forecasting.
\newblock {\em Available at SSRN 4630419}, 2023.

\bibitem{gentinetta2024complexity}
Gian Gentinetta, Arne Thomsen, David Sutter, and Stefan Woerner.
\newblock The complexity of quantum support vector machines.
\newblock {\em Quantum}, 8:1225, 2024.

\bibitem{yin2024experimental}
Zhenghao Yin, Iris Agresti, Giovanni de~Felice, Douglas Brown, Alexis Toumi, Ciro Pentangelo, Simone Piacentini, Andrea Crespi, Francesco Ceccarelli, Roberto Osellame, et~al.
\newblock Experimental quantum-enhanced kernels on a photonic processor.
\newblock {\em arXiv preprint arXiv:2407.20364}, 2024.

\bibitem{ding2025quantum}
Chao Ding, Shi Wang, Yaonan Wang, and Weibo Gao.
\newblock Quantum machine learning for multiclass classification beyond kernel methods.
\newblock {\em Physical Review A}, 111(6):062410, 2025.

\bibitem{schuld2019quantum}
Maria Schuld and Nathan Killoran.
\newblock Quantum machine learning in feature hilbert spaces.
\newblock {\em Physical review letters}, 122(4):040504, 2019.

\bibitem{rath2024quantum}
Minati Rath and Hema Date.
\newblock Quantum data encoding: A comparative analysis of classical-to-quantum mapping techniques and their impact on machine learning accuracy.
\newblock {\em EPJ Quantum Technology}, 11(1):72, 2024.

\bibitem{gonzalez2024efficient}
Javier Gonzalez-Conde, Thomas~W Watts, Pablo Rodriguez-Grasa, and Mikel Sanz.
\newblock Efficient quantum amplitude encoding of polynomial functions.
\newblock {\em Quantum}, 8:1297, 2024.

\bibitem{di2021improving}
Olivia Di~Matteo, Anna McCoy, Peter Gysbers, Takayuki Miyagi, RM~Woloshyn, and Petr Navr{\'a}til.
\newblock Improving hamiltonian encodings with the gray code.
\newblock {\em Physical Review A}, 103(4):042405, 2021.

\bibitem{larocca2025barren}
Martin Larocca, Supanut Thanasilp, Samson Wang, Kunal Sharma, Jacob Biamonte, Patrick~J Coles, Lukasz Cincio, Jarrod~R McClean, Zo{\"e} Holmes, and Marco Cerezo.
\newblock Barren plateaus in variational quantum computing.
\newblock {\em Nature Reviews Physics}, pages 1--16, 2025.

\bibitem{calabrese2004entanglement}
Pasquale Calabrese and John Cardy.
\newblock Entanglement entropy and quantum field theory.
\newblock {\em Journal of statistical mechanics: theory and experiment}, 2004(06):P06002, 2004.

\bibitem{leone2024practical}
Lorenzo Leone, Salvatore~FE Oliviero, Lukasz Cincio, and Marco Cerezo.
\newblock On the practical usefulness of the hardware efficient ansatz.
\newblock {\em Quantum}, 8:1395, 2024.

\bibitem{yang2021medmnist}
Jiancheng Yang, Rui Shi, and Bingbing Ni.
\newblock Medmnist classification decathlon: A lightweight automl benchmark for medical image analysis.
\newblock In {\em 2021 IEEE 18th International Symposium on Biomedical Imaging (ISBI)}, pages 191--195. IEEE, 2021.

\end{thebibliography}

\end{document}